\documentstyle[twocolumn,aps,floats,psfig]{revtex}

\begin{document}
\draft                   

\title{
       Thermodynamics of finite magnetic two-level systems }

\author{
   Peter Borrmann, Heinrich Stamerjohanns, Eberhard Hilf }

\address{
   Department of Physics of the University Oldenburg, 
   D-26111 Oldenburg, Germany }

\author{
   Philippe Jund, Seong Gon Kim, and David Tom\'anek }

\address{
   Department of Physics and Astronomy and
   Center for Fundamental Materials Research, \\
   Michigan State University, East Lansing, Michigan 48824-1116,
   USA }


\maketitle

\begin{abstract}
We use Monte Carlo simulations to investigate the thermodynamical
behaviour of aggregates consisting of few superparamagnetic particles
in a colloidal suspension. The potential energy surface of this
classical two-level system with a stable and a metastable ``ring''
and ``chain'' configuration is tunable by an external magnetic field
and temperature. We determine the complex ``phase diagram'' of this
intriguing system and analyze thermodynamically the nature of the
transition between the ring and the chain ``phase''.
\end{abstract}
With progressing miniaturization of devices, there is growing
interest in the thermodynamical behaviour of finite-size systems
\cite{cluster-thermodynamics}. A central question in this respect is,
whether small systems can exhibit well-defined transitions that could
be interpreted as a signature of phase transitions which, strictly
speaking, are well defined only in infinite systems. So far,
reproducible features of the specific heat have been interpreted as
indicators of ``melting'' transitions in small rare gas clusters
\cite{Berry,borr1}. 

Here we investigate the intriguing thermodynamical behaviour of a
structurally relaxed finite system which is controlled by {\em two}
external variables, namely the temperature $T$ and the magnetic
field $B_{\rm ext}$. The system of interest consists of few
near-spherical, superparamagnetic particles 
with a diameter of $\approx 10-500$~{\AA} in a colloidal suspension. 
Such systems, covered
by a thin surfactant layer, are readily available in macroscopic
quantities, are called ferrofluids, and are known 
to form complex labyrinthine \cite{Dick93} or
branched structures \cite{Wang94} as many particle systems, 
whereas the only stable states for
systems with few particles ($N<14$) are the ``ring'' and the ``chain''
\cite{PRLxFFR}.

The existence of two environmental variables,
yet still only two isomer states, gives rise to a
thermodynamical behaviour of unprecedented richness, as compared to
that of other small clusters, such as the noble gas clusters
\cite{Berry,borr1}. This is also an intriguing example of a
classical, externally tunable finite two-level system.

We will show that the system exhibits phase transitions between
{\em two} ordered phases, one magnetic and the other one nonmagnetic,
as well as phase transitions between these ordered phases and a
disordered phase. Whereas the system is not susceptible to small
magnetic fields, it shows a strong paramagnetic response when exposed
to larger magnetic fields.

Our model system consists of six spherical magnetite particles with a
diameter of $\sigma = 200$~{\AA}, mass $m=1.31{\times}10^7$~amu, and a
large permanent magnetic moment $\mu = 1.68{\times}10^5~\mu_{\rm B}$.
The potential energy $E_{\rm p}$ of this system in the external
field $\vec{B}_{\rm ext}$ consists of the interaction between each
particle $i$ and the applied field, given by $u_i=-\vec{\mu}_i \cdot
\vec{B}_{\rm ext}$, and the pair-wise interaction between the
particles $i$ and $j$, given by \cite{PRLxFFR}
\begin{eqnarray}
u_{ij}& = &(\mu_0^2/r_{ij}^3)
         \left[ \hat{\mu}_i\cdot\hat{\mu}_j
            - 3(\hat{\mu}_i\cdot\hat{r}_{ij})
               (\hat{\mu}_j\cdot\hat{r}_{ij})
         \right]  \nonumber \\
      & + &\epsilon
         \left[ \exp\left(-\frac{r_{ij}-\sigma}{\rho} \right)
              - \exp\left(-\frac{r_{ij}-\sigma}{2 \rho}\right)
        \right] \; .
\label{Eq1}
\end{eqnarray}
The first term in Eq.~(\ref{Eq1}) is the magnetic dipole-dipole
interaction energy. The second term describes a
non-magnetic interaction between the surfactant covered tops in a
ferrofluid that is repulsive at short range and attractive at long
range \cite{Wang94}. We note that the most significant part of this
interaction, which we describe by a Morse-type potential with
parameters $\epsilon=0.121$~eV and $\rho=2.5$~{\AA}, 
is the short-range repulsion, since even at
equilibrium distance the attractive part does not exceed $10\%$ of
the dipole-dipole attraction.

The equilibrium structures of small clusters are either rings or
chains, both of which can be easily distinguished by their different
mean magnetic moment $\langle \mu \rangle$. 

In the following, we present the first complete ``phase diagram'' of
a model aggregate of magnetic tops and describe in detail how ``phase
transitions'' occur in such a nanoscale system. Unlike in the bulk,
where transitions between well-defined phases are sharp, small
aggregates in a ferrofluid transform smoothly from one configuration
to another due to changes in the two environmental variables $B_{\rm
ext}$ and $T$. 
Our results presented below are based on a careful analysis  of 
set of 32  extensive Metropolis Monte Carlo simulations 
\cite{Metropolis}, each of which consisting of $6{\times}10^9$ steps. 

The canonical partition function, from which all thermodynamical
quantities can be derived, is given by
\begin{eqnarray}
Z(B_{\rm ext},T) &= & (2 \pi \beta)^{-6 N/2} \int
    \left[ \prod_{i=1}^{N} {\rm d}\vec{x}_{i}\;{\rm d}\phi_{i}\;
    {\rm d}\theta_{i}\;{\rm d}\psi_{i} \right] \nonumber \\
 &\times&    \exp\left( -\beta (\sum_{i<j}^{N} u_{ij}
    - \sum_i^N \mu_{i,\rm z} B_{\rm ext} )
    \right)  \; ,
  \label{Eq2}
\end{eqnarray}
where $\beta=(k_{\rm B}T)^{-1}$ and where the field $\vec{B}_{\rm
ext}$ is aligned with the $z$-axis. The pre-exponential factor
addresses the fact that each particle has three rotational and three
center-of-mass degrees of freedom. The key quantities we monitor
as a function of $T$ and $B_{\rm ext}$ are the formation enthalpy
$E^*$ of the isolated system that is decoupled from the
external field, given by $E^* = \sum_{i<j} u_{ij} = E_{\rm p} + \mu_{z}
B_{\rm ext}$, and the $z$-component of the total magnetic moment of
the aggregate, $\mu_{z}$. By monitoring these two response quantities
to the external temperature and field during all simulations, we
determine the weighted density of states
\begin{figure*}[htb]
\centerline{\psfig{figure=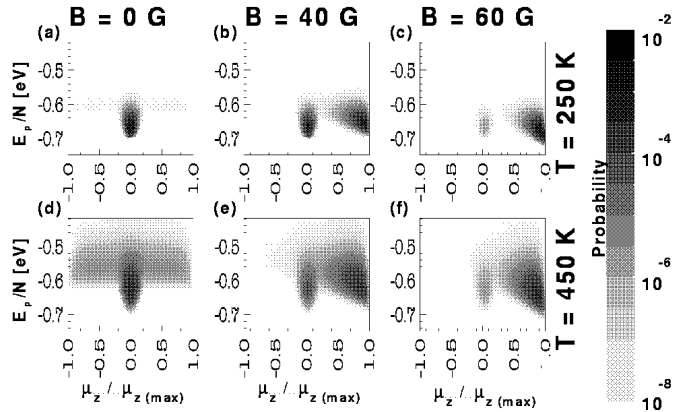,height=3.0in,angle=270}}
\caption{
Monte Carlo results for the probability to find an aggregate in a
state with its magnetic moment in the field direction $\mu_{z}$ and
potential energy $E_{\rm p}$. The individual contour plots show our
results for the temperature $T=250$~K at the field values
(a) $B_{\rm ext}$=0~G,
(b) $B_{\rm ext}$=40~G,
(c) $B_{\rm ext}$=60~G,
and $T=$450~K at the field values
(d) $B_{\rm ext}$=0~G,
(e) $B_{\rm ext}$=40~G,
(f) $B_{\rm ext}$=60~G.}
\label{Fig1}
\end{figure*}
\begin{eqnarray}
P(E^*,\mu_z;B_{\rm ext},T)
  & =&  Z^{-1}(B_{\rm ext},T) \; \rho(E^*,\mu_z) \;  \nonumber \\
 & \times &     \exp(-\beta (E^* -\mu_z B_{\rm ext})) \; ,
\end{eqnarray}
which gives the probability to find the system in a state with the
formation enthalpy $E^*$ and magnetic moment $\mu_{z}$. The
partition function $Z$, which appears as the normalization constant,
can be rewritten as
\begin{eqnarray}
Z(B_{\rm ext},T)
   && =  (2\pi\beta)^{-6 N/2}  \\
   &&  \int {\rm d}E^* {\rm d}\mu_z \;
      \rho(E^*,\mu_z)\; \exp(-\beta (E^*-\mu_z\; B_{\rm ext})) \;. 
       \nonumber
\end{eqnarray}
We then combined the results of all simulations using the
optimized Monte Carlo data analysis of Ferrenberg et al.
\cite{Ferren88,Ferren89} in order to calculate the normalized density
of states ${\rho}(E^*,\mu_z)$  and ${\rho}(E_p,\mu_z)$
with minimized statistical error
\cite{Ferren-2d}. Using the above defined density of states, the
field- and temperature dependence of the expectation value of any
function $F(E^*, \mu_z)$ can be obtained as
\begin{eqnarray}
\label{expec}
&&< F(E^*, \mu_z; B_{\rm ext},T) >
      =   \\
  &&   \int {\rm d}\mu_z\; \int {\rm d} E^* \;
        F(E^*,\mu_z) \; {P}(E^*,\mu_z;B_{\rm ext},T) \; .
    \nonumber
\end{eqnarray}
For the system described here we found it imperative to perform the
simulations at sufficiently high temperatures in order to cover the
whole configuration space properly. At low temperatures the thermal
equilibrium might not be achieved even after extremely long iteration
times, since transitions between rings and chains are very 
infrequent and might never occur.
In order to obtain a first idea about the stable and metastable states
of the system, we plotted in Fig.~\ref{Fig1} the probability
$\tilde{P}(E_{\rm p},\mu_z;B_{\rm ext},T)$ of finding the aggregate in a
state with potential energy $E_{\rm p}$ and total magnetic moment in the
field direction $\mu_z$. This is the projection of the probability to
find the system in a specific state in the high-dimensional
configuration space onto the $(E_{\rm p},\mu_z)$ subspace. Regions in the
$(E_{\rm p},\mu_z)$ subspace with a high probability indicate not only the
energetic preference of corresponding states, but also a large
associated phase space volume.

Rings always have an absolute magnetic moment $|\mu/\mu^{\rm max}|$
that is close to zero. Consequently, also the $z$-component of the
magnetic moment of rings is near zero, as seen in Fig.~\ref{Fig1}.
Even though the absolute magnetic moment $|\mu/\mu^{\rm max}|$ of
chains is close to one, these aggregates can not easily be
distinguished from rings in the absence of a field. In zero field, chains
have no orientational preference and the $z$-component of their
magnetic moment, $\mu_z/\mu_z^{\rm max}$, also averages to zero. Of
course, chains -- unlike rings -- do align with a nonzero magnetic
field and, especially at low temperatures, show a magnetic moment
$\mu_z/\mu_z^{\rm max}{\approx}1$ in the field direction.

The relative stability of an aggregate is reflected in its potential
energy $E_{\rm p}$. We find $E_{\rm p}$ to increase (corresponding to
decreasing stability) with increasing temperature. On the other hand,
applying a magnetic field destabilizes rings in favour of 
field-aligned chains. With increasing field, chains are confined to a
gradually decreasing fraction of the configurational space which
sharpens their distribution in the $(E_{\rm p},\mu_z)$ subspace, as seen
when comparing Figs.~\ref{Fig1}(a)--(c) and Figs.~\ref{Fig1}(d)--(f).

Under all conditions, we find two more or less pronounced local maxima
in the probability distribution $P$, corresponding to a ring with
$0{\alt}\mu_z/\mu_z^{\rm max}{\ll}1$, and a chain with
$0{\ll}\mu_z/\mu_z^{\rm max}{\alt}1$. At zero field
we observe a predominant occupation of
the more stable ring state. Due to the relatively small energy
difference with respect to the less favourable chain
${\Delta}E_{\rm p}^{\rm cr} /N = (E_{\rm p}^{\rm chain}-
E_{\rm p}^{\rm ring})/N =
0.06$~eV, both states get more evenly occupied at higher temperatures.
At fields as low as $B_{\rm ext}=40$~G, the energy difference
between chains and rings drops significantly to
${\Delta}E_{\rm p}^{\rm cr} /N = 0.02$~eV. As seen in Fig.~\ref{Fig1}(b),
this results in an equal occupation of both states even at low temperatures.
At the much higher field value $B_{\rm ext}=60$~G, chains are
favoured with respect to the rings by a considerable amount of energy
${\Delta}E_{\rm p}^{\rm cr}/N = -0.2$~eV. This strongly suppresses the
occurrence of rings, as seen in Figs.~\ref{Fig1}(c) and (f).

Fig.~\ref{Fig1} shows not only the stable and metastable states under
given conditions, but also the states found along the preferential
transition pathway between a ring and a chain in the projected
$(E_{\rm p},\mu_z)$ subspace. During the transition between a chain and a
ring, each aggregate must undergo a {\em continuous} change of $E_{\rm p}$
and $\mu_z$. The favoured transition pathways are then associated with
high-probability trajectories in the $(E_{\rm p},\mu_z)$ subspace. The
value of the activation barrier ${\Delta}E_{\rm p}^{\rm act}$ is then
given by the largest increase of $E_{\rm p}$ along the optimum transition
path which connects the stable and metastable ring and chain islands.
In our simulations we found that the activation barrier occurred
always at $\mu_z/\mu_z^{\em max}{\approx}0.22$. Consequently, we
concluded that the field dependence of the activation energy follows
the expression ${\Delta}E_{\rm p}^{\rm act} (B_{\rm ext}) =
{\Delta}E_{\rm p}^{\rm act}(B_{\rm ext}=0) - 0.22 \; \mu_{z}^{\rm max}\;
B_{\rm ext}$.

\begin{figure}
\centerline{\psfig{figure=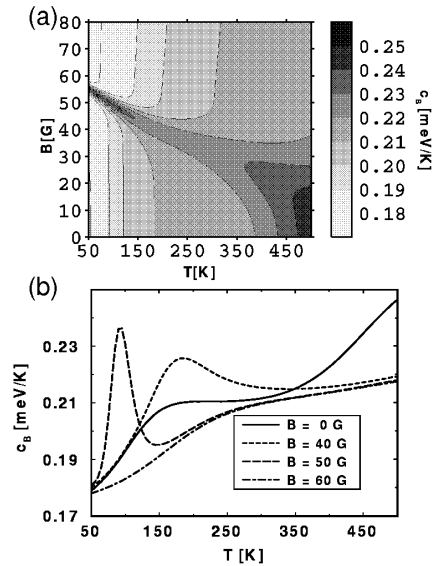,height=3.2in}}
\caption{
Specific heat per particle $c_B$ of the system as a function of
temperature $T$ and the external magnetic field $B_{\rm ext}$.
Results for the entire temperature and field range investigated here
are presented as a contour plot in (a). The temperature dependence of
$c_B$ for selected values of $B_{\rm ext}$ is presented in (b). 
}
\label{Fig2}
\end{figure}

In order to quantitatively describe the ``phase transitions'' occurring
in this system, we
focused our attention on the specific heat and the magnetic
susceptibility. The specific heat per particle
in a canonical ensemble is given by $c_{B} = {\rm
d}{\langle}E/N{\rangle}/{\rm d}T$, where the total energy is given by
$E = \frac{3}{2} N k_{\rm B} T + E_{\rm p}$. Correspondingly, we define
the magnetic susceptibility per particle
as $\chi = {\rm d}{\langle}\mu_z/N{\rangle}/{\rm d}B_{\rm ext}$.
These response functions are related to the
fluctuations of $E_{\rm p}$ and $\mu_z$ by
\begin{eqnarray}
c_{B} & = &  \left[ \frac{6N}{2} k_{\rm B}
            + k_{\rm B} \beta^2 (\langle E^{2}\rangle
                                -\langle E\rangle^{2}) \right]/N\;,\\
\chi  & = & \left[ \beta (\langle\mu_z^2\rangle
                   - \langle\mu_z \rangle^{2}) \right]/N\; .
\end{eqnarray}
Phase transitions are only well-defined in infinite systems and 
are associated with a discontinuous change in the total
energy and specific heat when crossing the phase boundary. The
corresponding changes in finite systems are more gradual. This is
seen in the well defined, yet not sharp ``crest line''
separating the ring and the chain ``phase'' in the $T-B_{\rm ext}$
``phase diagram'' in Fig.~\ref{Fig2}(a). These results illustrate
how the critical magnetic field for the ring-chain
transition decreases with increasing temperature. At high
temperatures, the ``line'' separating the ``phases'' broadens
significantly into a region where rings and chains coexist.
\begin{figure}
\centerline{\psfig{figure=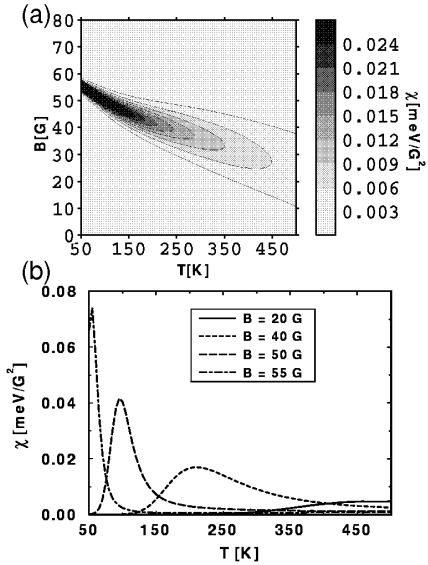,height=3.2in}}
\caption{
Magnetic susceptibility per particle $\chi$
of the system as a function of temperature $T$ and the external
magnetic field $B_{\rm ext}$. Results for the entire temperature and
field range investigated here are presented as a contour plot in (a).
The temperature dependence of $\chi$ for selected values of $B_{\rm
ext}$ is presented in (b). 
}
\label{Fig3}
\end{figure}

The line plot in Fig.~\ref{Fig2}(b) is the respective
constant-field cut through the contour plot
in Fig.~\ref{Fig2}(a). As can be seen in Fig.~\ref{Fig2}(b), there is
no transition from chains to rings, indicated by a peak in $c_{B}$,
at fields exceeding $50$~G, which is close to the critical
field value at which chains become favoured over rings at zero
temperature. At fields $B_{\rm ext}<<40$~G, on the other hand,
there is no region where chains would be thermodynamically preferred
over the rings, and we only observe a gradual transition from the ring
phase into the coexistence region with increasing temperature. The
specific heat behaviour at zero field resembles that of a small system
with a gradual {\em melting} transition close to $150$~K and an onset
of disorder at about $350$~K \cite{Berry83}. 
As seen in Fig.~\ref{Fig2}(b), the critical temperature and the width
of the transition region can be externally tuned by the second
thermodynamical variable, the external magnetic field $B_{\rm ext}$.

Fig.~\ref{Fig3} displays the magnetic susceptibility $\chi$, another
prominent indicator of phase transitions in magnetic systems, as a
function of T and $B_{\rm ext}$. Like the specific heat in
Fig.~\ref{Fig2}(a), the crest line in $\chi$ separates the chain
``phase'' from the ring ``phase'' in this $T-B_{\rm ext}$ ``phase
diagram''. Moreover, Fig.~\ref{Fig3}  reveals
the fundamentally different magnetic character of these ``phases''.
Whereas the system is nonmagnetic in the ring ``phase'' found below
$40$~G, it behaves like a ferromagnet consisting of Langevin
paramagnets in the chain ``phase'' at higher fields. The transition
between these states is again gradual, as expected for finite
systems. The line plot in Fig.~\ref{Fig3}(b) is the
respective constant-field cut through the
contour plot in Fig.~\ref{Fig3}(a). 
When the system is in the chain ``phase''
it behaves like a paramagnet obeying the Curie-Weiss law, as can be
seen in Fig.~\ref{Fig3}(b)\cite{LowTempLimit}.

At relatively low temperatures, where the aggregates are intact, the
expectation value of the magnetic moment 
first increases due to the gradual conversion from
nonmagnetic rings to paramagnetic chains. According to
Fig.~\ref{Fig3}(b), this uncommon behaviour persists up to $T = 200$~K
at B$_{\rm ext} = 40$~G. This trend is reversed at higher
temperatures, where all aggregates eventually fragment into single
paramagnetic tops. In this temperature range, the magnetic moment as
well as the susceptibility decreases with increasing
temperature.

Since the transition probability between both states
is extremely low at low temperatures and fields, magnetically
distinguishable metastable states can be frozen in. A chain
configuration, which is metastable in zero field, can be prepared by
first annealing the system to $T{\agt}350$~K and subsequent quenching
in a strong field. Similarly, a frozen-in ring configuration is
unlikely to transform to a chain at low temperatures, unless exposed
to very large fields. Thus the above described phase diagrams can be used
to externally manipulate the self-assembly of magnetic nanostructures.

In conclusion we have studied the thermodynamical behaviour of a
finite two-level system, which is externally tunable by two
independent variables, namely the temperature and the magnetic field.
Much of the behaviour encountered in this system, such as transitions
between different states, has a well-defined counterpart in infinite
systems. The reason for the encountered richness of the thermodynamic
and magnetic properties is the relative ease of structural
transformations, which is typical for finite systems. Consequently, we
expect other finite magnetic systems, e.g. 
small transition metal clusters, where a small number of structural
isomers with substantially different magnetic moments could coexist
\cite{OLMSU1}, to follow this behaviour. 
Moreover, we expect that our results can also be transferred to 
nanocrystalline material, such as magnetic clusters
encapsulated in the supercages of zeolites, which will likely retain
some of the intriguing properties of the isolated finite systems.\\

DT, PJ and SGK acknowledge financial support by the 
NSF under Grant No. PHY-92-24745 and the 
ONR under Grant No. N00014-90-J-1396.


\begin{references}
\bibitem{Ferren-2d} We extended the Ferrenberg analysis
        in a straight-forward way to deal with
        a two-dimensional density of states.
\bibitem{LowTempLimit} Our numerical approach did not allow to
     investigate the temperature region below $50$~K. The onset of the
     Curie-Weiss behaviour, indicated by a maximum in $\chi$ at nonzero
     temperature,
     can be seen whenever the system is in the chain ``phase''.
\end{references}
\end{document}